%% file: euroloop.tex
\documentstyle[epsfig,latexsym]{europhys}
\input EuroMacr

\begin{document}
\euro{}{}{}{} \Date{} \shorttitle{L. PELITI: EVOLUTIONARY LOOP}
\title{%
A solvable model of the evolutionary loop}
\author{Luca
  Peliti\inst{1,2}} 
\institute{
  \inst{1}Laboratoire de Physico-Chimie Th\'eorique, CNRS URA 1382,\\
  ESPCI, 10, rue Vauquelin, F-75231 Paris Cedex 05 (France).\\
  \inst{2}Dipartimento di Fisica and Unit\`a INFM, Universit\`a
  ``Federico II",\\
  Mostra d'Oltremare, Pad.~19, I-80125 Napoli
  (Italy).\footnote{Permanent  address.}\\
  Associato INFN, Sezione di Napoli. E-mail: {\tt peliti@na.infn.it}}
\pacs{ 
  \Pacs{02}{50.Ey}{Stochastic processes}
  \Pacs{05}{40.+j}{Fluctuation phenomena, random processes, and
    Brownian motion} 
  \Pacs{87}{10.+e}{General, theoretical and
    mathematical biophysics} } 
\maketitle
\begin{abstract}
  A model for the evolution of a finite population in a rugged fitness
  landscape is introduced and solved. The population is trapped in an
  evolutionary loop, alternating periods of stasis to periods in which
  it performs adaptive walks. The dependence of the average rarity of
  the population (a quantity related to the fitness of the most
  adapted individual) and of the duration of stases on population size
  and mutation rate is calculated.
\end{abstract}
The simplest conceivable evolutionary situation is a
population of asexually reproducing individuals set in a fixed
environment.  The reproductive power of an individual is measured
by its fitness, {\it i.e.}, by a quantity proportional to the expected
number of its offspring~\cite{Maynard-Smith,Peliti}. In model
building, one often assumes that the fitness is determined by the
genotype, and that the genotype itself is transmitted identically from
parent to offspring, apart from mutations.

The role of mutations is favorable if the overall fitness of the
members of the population is low, because they allow the population to find
genotypes with higher fitness, {\it i.e.}, to adapt.
This can be represented, following Sewall Wright~\cite{Sewall},
by saying that the population approaches a fitness peak.
On the other hand, mutations become pernicious when the population is
located on such a peak, since they may let the population
lose contact with it.
In fact, the two effects have different relevance depending
on population size: if the population size is large, and the
fitness small, adaptation dominates; but if the population is small,
and the fitness peak, no matter how lofty, is narrow,
mutations have a negative effect.

This situation has been described by two classes of
models:
\begin{enumerate}
\item In the quasispecies model~\cite{Eigen} one takes the infinite population
size limit from the outset, obtaining an equation (akin to a Master Equation)
for the genotype distribution in the population. It is interesting that,
if the fitness peaks are narrow enough, this equation exhibits a transition
(the {\it error threshold\/}) between an adaptive regime, and a regime
in which adaptation is irrelevant. Nevertheless, the description of the
non-adaptive regime is not satisfactory within this class of models.
\item The adaptation process has been described as a special kind of
random walk, the {\it adaptive walk}, by Kauffman and Levine~\cite{Kauffman}
and others. In this model, the fitness can only increase, and mutations
only have positive effects. An ``annealed" version of this model has
been exactly solved by Flyvbjerg and Lautrup~\cite{Flyvbjerg}.
\end{enumerate}
The {\it stochastic escape\/} of a finite population from a narrow fitness
peak has been discussed by Higgs and Woodcock~\cite{Higgs1}.
They find that, in the same limit in which the error threshold appears in its
fullest glory in the quasispecies model, a finite population eventually
loses contact with the adaptation peak. Building on this observation
and on numerical simulations, Woodcock and Higgs~\cite{Higgs} have
been led to describe the behavior of a population evolving in a rugged
fitness landscape ({\it i.e.}, in a situation where even slight
changes of the genotype lead to arbitrarily large changes in the
fitness) as an {\it evolutionary loop\/}:
\begin{enumerate}
\item If the fitness of the population is low, favorable mutations
get fixed in the population, which thus performs an adaptive walk,
reaching a local fitness peak.
\item The population can be evicted from the adaptation peak by
stochastic escape, and start a new adaptive walk from a random,
usually low, fitness value.
\end{enumerate}

In this Letter, I introduce a solvable model that exhibits such a
behavior.  The model is a slight generalization of the Annealed
Adaptive Walk Model introduced and solved by Flyvbjerg and
Lautrup~\cite{Flyvbjerg}, and allows for stochastic escape. It depends
on only two parameters, namely population size and mutation rate. I
argue that it should describe any mutation-selection model in the
strong selection limit, {\it i.e.}, when the fitness distribution is
broad, provided that the correct variables are identified and the
correct scaling of the parameters is performed.  This conclusion is
borne out by simulations of a slightly more realistic model which I
report at the end of the Letter.

I consider a population of $M$ individuals ($M$ is fixed) evolving in
a rugged fitness landscape. At each generation, for each member
$\alpha$ of the new population, its parent $\alpha'$ is chosen among
the old population. The probability $W_\gamma$ that the individual
$\gamma$ is chosen to reproduce is given by
\begin{equation}
W_{\gamma}=\frac{F_{\gamma}}{\sum_\beta F_\beta},\label{repro}
\end{equation}
where $F_\gamma$ is the fitness of the individual $\gamma$. The
genotype of $\alpha$ is taken to be equal to that of its parent
$\alpha'$ apart from mutations. I denote by $u$ the probability that
the offspring undergoes at least one mutation.

I make the following simplifications:
\begin{description}
\item[Rugged fitness landscape:] The fitnesses $F$ of different
genotypes are independent random variables, identically distributed
according to some distribution $\rho(F)$;
\item[Infinite genome limit:] On each mutation,
 a wholly new genotype appears in
the population;
\item[Strong selection limit:] At each generation, only the fittest
individuals are allowed to reproduce.
\end{description}
Given a fitness distribution $\rho(F)$, one can
define the {\it rarity\/} $h$ of a fitness
value $F$ as the probability that, picking a new
fitness at random with the distribution $\rho(F)$ one obtains
a higher fitness value~\cite{Higgs}:
\begin{equation}
h(F)=\int_F^\infty\drm F'\,\rho(F').
\end{equation}
Note that higher fitness corresponds to smaller rarity.
Within our hypotheses, it is easy to see that the genetic properties of the
whole population are summarized by the rarity
of its fittest individual, which we shall simply call
the rarity of the population.
Consider the population at generation $t$, and let $h(t)$ be its rarity. 
For each individual of the new generation,
decide (with probability $u$) if it mutates. If it does,
assign to it a rarity $h_\alpha$ with uniform probability between 0 and 1.
Otherwise, assign to it a rarity $h_\alpha$ equal to $h$.
The rarity of the population is then
\(
h(t+1)=\min_\alpha h_\alpha.
\)
Therefore, the evolution of an {\it ensemble\/} of populations
is described by the probability distribution function $P(h,t)$
of their rarity.

In the following I derive the evolution equation for $P(h,t)$ and solve
for its asymptotic form in the limit of large, but finite populations.

Let us denote by $\pi^{M}_{\nu}$ the probability that, at any one generation,
exactly $\nu$ individuals mutate:
\begin{equation}
\pi^M_\nu={M\choose\nu}u^\nu (1-u)^{M-\nu},
\end{equation}
and by $\phi_\nu(h)$ the probability that, picking up $\nu$ genotypes
at random, one obtains rarities all larger than $h$:
\begin{equation}
\phi_\nu(h)=(1-h)^\nu.
\end{equation}
It is then easy to see that $P(h,t)$ satisfies the following equation:
\begin{eqnarray}
P(h,t+1)&=&\pi^M_0P(h,t)-\pi^M_M\phi'_M(h)\nonumber\\
&&{}+\sum_{\nu=1}^{M-1}\pi^M_\nu\left[\phi_\nu(h)P(h,t)-\phi'_\nu(h)
\int_h^1\drm h'\,P(h',t)
\right].\label{evol}
\end{eqnarray}
The first term represents the case in which there are
no mutations, and the second one the case in which all
individuals mutate, and the rarity of the new
population is equal to $h$. The sum represents the
case in which some individuals mutate, and some do  not:
the first term in square brackets describes the case
where the rarity of all mutated individuals exceeds $h$, and the
second one the case in which the rarity of at least one
mutated individual is equal to $h$ and smaller than the
preceding rarity of the population.
Introducing the notations
\begin{eqnarray}
\Phi(h,t)&=&\int_h^1\drm h'\,P(h',t),\\
f(h)&=&\sum_{\nu=0}^{M-1}\pi^M_\nu\phi_\nu(h)=(1-uh)^M-u^M(1-h)^M,\\
g(h)&=&-\pi^M_M\phi'_M(h)=u^M M(1-h)^{M-1},
\end{eqnarray}
eq.~(\ref{evol}) can be written
\begin{equation}
\Phi'(h,t+1)=f(h)\Phi'(h,t)+f'(h)\Phi(h,t)-g(h)=\dif{}{h}
\left[f(h)\Phi(h,t)\right]-g(h),
\end{equation}
where the primes denote derivatives with respect to $h$.
The solution of this equation is
\begin{equation}
\Phi(h,t)=\left(f(h)\right)^t\Phi(h,0)+\frac{1-\left(f(h)\right)^t}{
1-f(h)}\int_h^1\drm h'\,g(h').
\end{equation}
Let us consider large values of $M$ and set
\begin{equation}
h=\frac{ k }{M};\qquad 1-u=\frac{w}{M}.
\end{equation}
In this limit, one has
\begin{eqnarray}
f(h)&\simeq& e^{- k }-e^{-( k +w)};\\
\int_h^1\drm h'\,g(h')&\simeq& e^{-( k +w)}.
\end{eqnarray}
We obtain therefore 
\begin{equation}
\Psi( k ,t)\equiv \lim_{M\to\infty}\Phi( k /M,t)=
e^{- k  t}(1-e^{-w})^t
\Psi( k ,0)+\frac{1-e^{- k  t}(1-e^{-w})^t}%
{1-e^{- k }+e^{-( k +w)}}e^{-( k +w)}.
\end{equation}

Therefore, the $t\to\infty$ limit of $\Psi(k,t)$, 
denoted by $\Psi_\infty(k)$, is given by
\begin{equation}
\Psi_\infty( k )=\frac{1}{e^w(e^k-1)+1}.\label{solution}
\end{equation}
One has
\begin{equation}
\Psi_\infty(0)=1;\qquad \Psi_\infty( k )\sim e^{- k }\quad\mbox{for}\quad  k 
\to\infty.
\end{equation} It  
can be checked that the average value of $ k $
is given by
\begin{equation}
 \left< k \right>=-\int_0^\infty\drm k '\,\Psi'_\infty( k ') k '
=\frac{w}{e^w-1},\label{average:eq}
\end{equation}
and varies from $\left< k \right>=1$ to $\left< k \right>=0$
as $w$ varies between 0 and infinity.

Let us define {\it stasis\/} the regime in which the rarity $h$ of the
population remains constant or decreases, and {\it stochastic
  escape\/} the event by which it increases, since a harmful mutation
gets fixed in the population. We can then compute the distribution of
stasis duration as follows. Denote by $K( k ,t)$ the conditional
probability that, if the minimal rarity was equal to $ k /M$ at
generation $t=0$, it has never decreased up to generation $t$. This
function obeys the following equation:
\begin{eqnarray}
K( k ,t+1)&=&\left[1-\pi^M_M+\pi^M_M\phi_M( k /M)\right]K( k ,t)
\nonumber\\
&\simeq&\left[1-e^{-( k +w)}\right]K( k ,t),
\end{eqnarray}
from which we deduce
\begin{equation}
K( k ,t)=\left[1-e^{-( k +w)}\right]^t.
\end{equation}
The average duration of a stasis is therefore
\begin{equation}
\left<t\right>=e^{ k +w}-1.
\end{equation}
If we take as a ``typical" value of $ k $ that for which
$\Psi_\infty( k )=\frac{1}{2}$, we obtain $\left<t\right>=e^w$.

\begin{figure}[htb]
\begin{center}
\epsfig{file=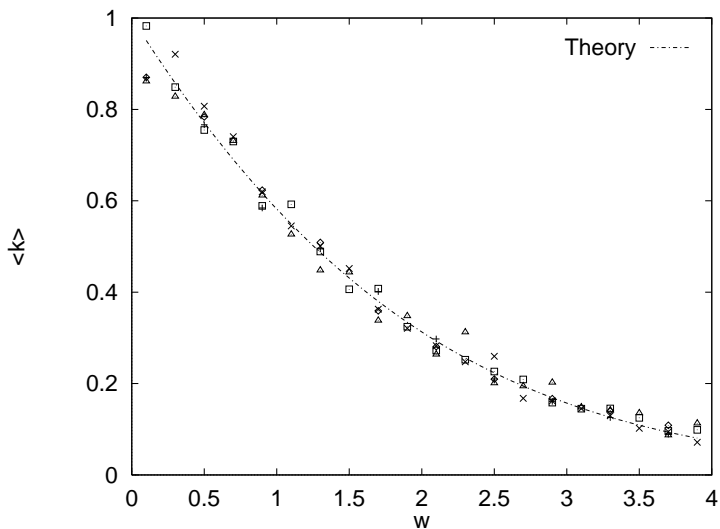,width=10cm}\end{center}
\caption{Average scaled rarity $\left<k\right>=M\left<h\right>$
  over $250\,000$ generations as a function of the scaled mutation
  rate $w=M(1-u)$. $\Diamond$: $M=32$, $s=0.1$; $+$: $M=32$, $s=0.2$;
  $\Box$: $M=128$, $s=0.1$; $\times$: $M=512$, $s=0.2$; $\triangle$:
  $M=512$, $s=0.4$. Root mean square deviations vary between 0.04 for
  $w\sim 0.1$ to 0.006 for $w\sim 4$. The dotted line corresponds to
  eq.~(\protect\ref{average:eq}).  }\label{average:fig}
\end{figure}

I now consider a population in which the probability $W_\alpha$ that
an individual $\alpha$ reproduces is given by eq.~(\ref{repro}).  For
the strong selection limit to hold, the probability that individuals
with less than the maximum fitness reproduce should be negligible.
This implies in turn that the fitness distribution function $\rho(F)$
should be very broad. I have therefore assumed that the fitnesses are
extracted from a L\'evy distribution $\rho(F)=s F^{-s-1}$ for $1\le
F<\infty$, which, for $0<s<1$, has an infinite second moment.  The
curves of $\left<k\right>$ {\it vs.} $w$ should be independent of $s$.
The results of the simulation this model are shown in
Fig.~\ref{average:fig}.  One notices that the points
tend to lie above the theoretical curve for larger values of $w$.
In fact, the deviations from the strong selection limit become
significant in this regime.

Indeed, the probability that the ratio of the largest to the second
largest fitness value, extracted from a L\'evy distribution of
parameter $s$, exceeds $\lambda$, is given by $\lambda^{-s}$,
independently of the number of extracted values. The strong selection
limit holds therefore only if among the individuals that
reproduce without mutations (on average  $w=(1-u)M$ at each generation), the
chance that the second fittest one reproduces is negligible. This
happens if the probability that this ratio is smaller than $w$ is
itself small, say of order $\epsilon$.  This implies in turn
$w^{-s}>1-\epsilon$, or, for small values of $s$, $w<e^{\epsilon/s}$.
It may easily be checked that this is indeed the case.

I am afraid that it would be difficult to observe stochastic escape in
all but the smallest populations, although the analogous effect in
smooth fitness landscapes (namely Muller's ratchet~\cite{Muller}) has
been observed during repeated genetic bottleneck transfers of
monoclonal antibody-resistant mutants of vesicular stomatitis virus,
or of mutants of an RNA bacteriophage~\cite{Chao,Duarte}.  I hope
nevertheless that this approach provides at least a better grasp of the
evolutionary behavior of a finite population in a rugged fitness
landscape, and may be used as a starting point for the study of
more realistic situations.

\stars I thank M. M\'ezard and J.-B. Fournier for useful suggestions.
I also thank A. Ajdari and G. Parisi for their interest in this work.
I acknowledge the support of a Chaire Joliot de l'ESPCI.

\end{document}

%% file: EuroMacr.tex
\def\And{{\rm and\ }}

\def\drm{{\rm d}}

\def\dif#1#2{\frac{\drm#1}{\drm#2}}

\def\stars{\bigskip\centerline{***}\medskip}

\newif\ifboo \boofalse

\def\Review#1{\boofalse{\it #1},}
\def\Name#1{{\sc #1},}
\def\Vol#1{\ifboo Vol. {\bf #1}\else{\bf #1}\fi}

\def\Book#1{\bootrue{\it #1},}
\def\Page#1{\ifboo {\rm p. #1}\else{\rm #1}\fi}

%% file: euroloop.bbl
\begin{thebibliography}{99}
  
\bibitem{Maynard-Smith}The concept of fitness is nicely discussed in
  \Name{J. Maynard Smith} \Book{Evolutionary Genetics} (Oxford U.P.,
  Oxford) 1989.
  
\bibitem{Peliti}An introduction to the statistical point of view in
  evolutionary theory and a recent bibliography can be found in
  \Name{L. Peliti} {\bf cond-mat}/9712027.
  
\bibitem{Sewall}\Name{S. Wright} \Review{Genetics} \Vol{16} (1931)
  \Page{97}.
  
\bibitem{Eigen}A comprehensive review of quasispecies models is
  contained in \Name{M. Eigen, J. Mc Caskill \And\ P. Schuster}
  \Review{Adv. Chem. Phys.}  \Vol{75} (1989) \Page{149}.
  
\bibitem{Kauffman}\Name{S. A. Kauffman \And\ S. Levine} \Review{J.
    theor. Biol.}  \Vol{128} (1987) \Page{11}.
  
\bibitem{Flyvbjerg}\Name{H. Flyvbjerg \And\ B. Lautrup} \Review{Phys.
    Rev. A} \Vol{46} (1992) \Page{6714}.
  
\bibitem{Higgs1}\Name{P. G. Higgs \And\ G. Woodcock} \Review{J. Math.
    Biol.} \Vol{33} (1995) 677.
  
\bibitem{Higgs}\Name{G. Woodcock \And P. G. Higgs} \Review{J. theor.
    Biol.} \Vol{179} (1996) \Page{61}.

\bibitem{Muller}\Name{H. J. Muller} \Review{Mutat. Res.} \Vol{1}
  (1964) \Page{2}.

\bibitem{Chao}\Name{L. Chao} in \Name{S. S. Morse} (ed.),
  \Book{The Evolutionary Biology of Viruses} (Raven, New York, 1994),
  \Page{233}.

\bibitem{Duarte}\Name{E. A. Duarte, D. Clarke, A. Moya, E. Domingo
\And\ J. Holland} \Review{Proc. Natl. Acad. Sci. Usa} \Vol{89} (1992) 6015;
\Name{E. A. Duarte, D. Clarke, S. F. Elena, A. Moya, E. Domingo
\And\ J. Holland} \Review{J. Virol.} \Vol{67} (1993) 3620.

\end{thebibliography}
